| | |
|---|---|
| **Source** | [1]University of British Columbia (UBC) & [2]TELUS Communications Inc. |
| **Status** | Input Document |
| **Title** | Quality Assessment of High Dynamic Range (HDR) Video Content Using Existing Full-Reference Metrics |
| **Author** | [1]Amin Banitalebi-Dehkordi, [1]Maryam Azimi, [1]Yuanyuan Dong, [1,2]Mahsa T. Pourazad, and [1]Panos Nasiopoulos |

## Abstract

The main focus of this document is to evaluate the performance of the existing LDR and HDR metrics on HDR video content which in turn will allow for a better understanding of how well each of these metrics work and if they can be applied in capturing, compressing, transmitting process of HDR data. To this end a series of subjective tests is performed to evaluate the quality of DML-HDR video database [1], when several different representing types of artifacts are present using a HDR display. Then, the correlation between the results from the existing LDR and HDR quality metrics and those from subjective tests is measured to determine the most effective exiting quality metric for HDR.

## 1 Introduction

The objective of this study is to evaluate the performance of the state-of-the-art full-reference quality metrics on HDR video content at the presence of different distortions. To this end, a new HDR video dataset is created, which consists of indoor and outdoor video sequences with different brightness, motion levels [1]. Five types of distortions (AWGN, intensity shifting, salt & pepper noise, low pass filtering, and compression) are applied to four HDR sequences of the DML-HDR dataset [1]. Then the quality of the distorted HDR videos are evaluated both subjectively and objectively.

One approach to evaluate the quality of HDR content objectively is to extend the usage of LDR quality metrics on HDR content. To this end the HDR data needs to be first processed so that its pixel value falls into a range that is supported by LDR quality metrics. This method is known as perceptually uniform (PU) encoding [2]. Another very simple but effective technique for employing LDR metrics on HDR data is based on the multi-exposure (ME) inverse tone mapping. In this technique the HDR stream is tonemapped to several LDR streams with different exposure range, and then the LDR metric is applied to each LDR stream and the numerical quality values are averaged at the end [3]. In addition to these LDR quality metric-based approaches, there are a limited number of quality metrics that have been developed specifically for HDR content. HDR-VDP-2 is known as one of the state-of-the-art HDR quality metrics that generates a quality value in addition to the distortion map [4]. The quality metrics used for objective assessments in this report are VIF, PSNR, and SSIM using both PU and ME methods, as well as HDR-VDP2. Then the correlation between the subjective and objective results is measured to evaluate the performance of the objective quality metrics in the presence of the tested distortions.

## 2 DML-HDR Dataset

"DML-HDR" dataset is captured by RED Scarlet-X professional cameras community [1]. Each video sequence in this dataset is approximately 10 seconds long with a frame rate of 30 frames per second (fps). All sequences are recorded in 2048×1080 resolution. Five video sequences of the DML-HDR dataset were selected for this study. Table I and Fig. 1 show the specifications and snapshots of the used test video sequences respectively. The captured videos are available both in RGBE and YUV 12-bit format. RGBE is a lossless HDR video format, where each pixel is encoded with 4 bytes, one byte for red mantissa, one byte for green mantissa, one for the blue mantissa, and one byte for a common exponent [1]. The YUV 12-bit format consists of three channels, Y for luma and U and V for Chroma. Each channel is represented by integer values between 0 and 4095 (12 bits).

TABLE I
DESCRIPTION OF THE HDR VIDEO DATASET

| Sequence Name | Motion Level | Number of Frames | Environment |
|---|---|---|---|
| Playground | Fast | 222 | Outdoor |
| Table | Slow | 261 | Indoor |
| Christmas | Intermediate | 317 | Indoor |
| Hallway | Intermediate | 253 | Indoor |

## 3 Experiment Setting

### A. Distortions

Five distinctive types of distortions are applied to each video sequence as listed below:
- ***Additive White Gaussian Noise (AWGN):*** white Gaussian noise with mean of zero and standard deviation of 0.002 was added to all frames of each video. Based on our knowledge from LDR videos, this value of standard deviation may seem to be too small. However, observations from watching distorted HDR videos on the HDR display showed that AWGN with the standard deviation value of 0.002 is visible. This may be due to their larger dynamic range compared to LDR videos. Note that, before adding the AWGN noise to the content, all pixel values were normalized between 0 and 1. After adding the AWGN noise, pixel values were converted back to the original scale.
- *Mean intensity shift:* the luminance of the HDR videos was globally increased in all the frames of each video sequence by 10% of the maximum scene luminance.
- *Salt and pepper noise:* Salt and pepper noise was added to the 2% of the pixels in each frame of the videos. The distribution of the affected pixels by salt and pepper noise was random.
- *Low pass filter*: An 8×8 Gaussian low pass filter with standard deviation of 8 was applied to each frame of all the sequences. Subsequently, rapid changes in intensity in each frame were averaged out.
- *Compression artifacts*: All the videos were encoded using the HEVC encoder (HM software version 12.1) with random access main10 profile configuration. The HEVC encoder settings were as follows: hierarchical B pictures, group of pictures (GOP) size of 8, internal bit-depth of 12, input video format of YUV 4:2:0 progressive, and enabled rate-distortion optimized quantization (RDOQ). The quantization parameter (QP) was set to 22, 27, 32, and 37 in order to simulate impaired videos with a wide range of compression distortions.

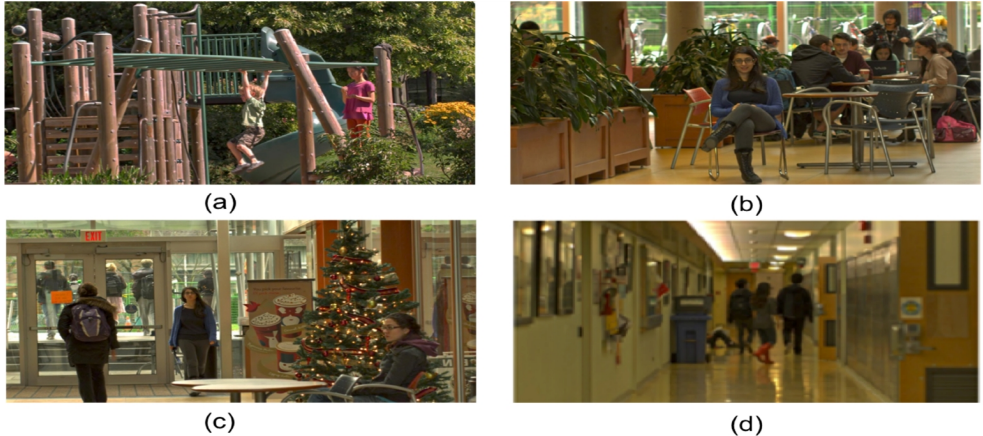

Fig 1. Snapshots of the first frames of HDR test video sequences (tone-mapped version): (a) Playground, (b) Table, (c) Christmas, and (d) Hallway

The compressed videos are available in 12-bit YUV format in the "DML-HDR" video dataset, whereas all other distorted videos are available in HDR format (.hdr).

## B. Objective quality metrics

In order to meaningfully use LDR metrics to evaluate the quality of HDR content, PU encoding and multi-exposure methods are employed. LDR metrics used in our experiment include PSNR, SSIM [5], and VIF [6]. Among the existing HDR metrics, HDR-VDP-2 is used in our experiment, as it is the state-of-the-art full-reference metric that works for all luminance conditions (both LDR and HDR) [4].

## C. Subjective test procedure

The videos were displayed on a Dolby HDR TV prototype built based on the concept explained in [7]. As illustrated in Fig. 2, this system consists of two main parts: 1) a 40 inch full HD LCD panel in the front, and 2) a projector with HD resolution at the back to provide the backside luminance. The contrast range of the projector is 2000:1. The original HDR video signal is split into two streams, which are sent to the projector and the LCD (see [7] for details). The input signal to the projector includes only the luminance information of the HDR content and the input signal to the LCD includes both luma and chroma information of the HDR video. The process for preparing the input signal to this system is as follows:
- Load HDR image in RGB space
- Tone-map HDR using Reinhard to generate RGB_LCD
- Extract luminance channel from HDR image: (Y = 0.2126 R + 0.7152 G + 0.0722 B), and normalize it to [0, 1]
- Evenly split pixel values between projector and LCD, Y_projector = $Y^{0.5}$
- Simulate point spread function of projector (Y_lightfield), low pass filter Y_projector by a Gaussian filter (window size: 12×12, sigma=2)
- LCD signal: RGB_LCD/ Y_lightfield

Using this configuration, the light output of each pixel is effectively the result of two modulations with the two individual dynamic ranges multiplied, yielding an HDR signal. This HDR display system is capable of emitting light at a maximum brightness level of 2700 cd/m$^2$.

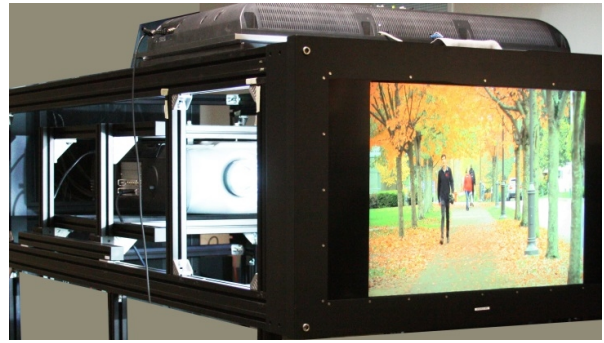

Fig 2. Prototype HDR Display

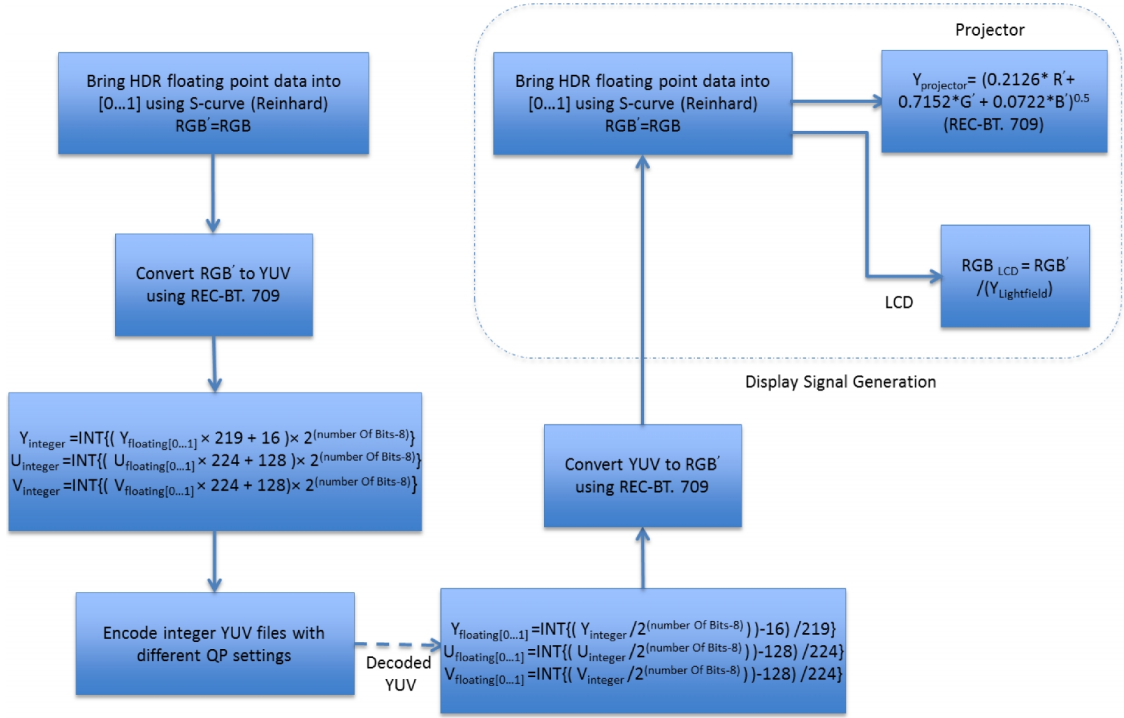

Fig. 3 Conversion of HDR data to YUV and to HDR display format

Fig. 3 shows the diagram for processing the content for viewing on this prototype display.

Prior to the actual experiment, a training session was shown to the observers to familiarize them with the rating procedure. The test sessions were designed based on the Double-Stimulus (DS) method [8]. In particular, after each 10-second long reference video, a 3-second gray interval was shown followed by the 10-second long distorted video. Another 4-second gray interval was allocated after the test video, allowing the viewers to rate the quality of the test video with respect to that of the reference one. The scoring is based on discrete scheme where a numerical value from 1 (worst quality) to 10 (identical quality) is assigned to each test video representing its quality with respect to the reference video [8] (Numerical Categorical Judgment Method). Note that in order to stabilize the subjects' opinion, a few dummy video pairs were presented at the beginning of the test and the subjects were asked to rate them. The collected scores for these videos were discarded from the final results. Eighteen adult subjects including 10 males and 8 females participated in our experiment. The subjects' age range was from 19 to 35 years old. Prior to the tests, all the subjects were screened for color blindness using the

Ishihara chart and visual acuity using the Snellen charts. Those subjects that failed the pre-screening test did not participate in the test.

## 4 Results

After collecting the subjective results, the outlier subjects were detected according to the ITU-R BT.500-13 recommendation in [8]. No outlier was detected in this test. The Mean Opinion Score (MOS) for each impaired video was calculated by averaging the scores over all the subjects with 95% confidence interval.

Fig. 4 shows the MOS versus birate curves for different video sequences. Table II summarizes the bitrate of the encoded videos at different QP settings. Fig. 5 shows the objective quality metric results versus subjective test results in the presence of compression artifacts.

Fig. 6 shows the objective quality metric results versus subjective test results in the presence of AWGN, Intensity shifting, salt & pepper noise, and low pass filtering impairments.

Fig. 7 shows the objective quality metric results versus subjective test results in the presence of all the impairments used in this study (i.e., AWGN, Intensity shifting, salt & pepper noise, low pass filtering, and compression).

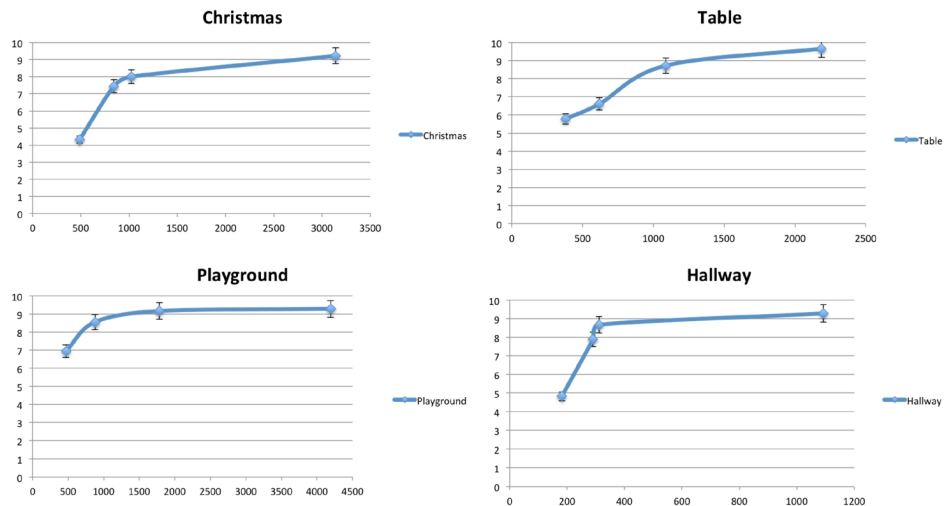

Fig.4 MOS – Rate plots of the encoded HDR video dataset

TABLE II
BIT-RATES OF ENCODED HDR VIDEOS

| Sequence Name | QP | Rate (Kb/sec) |
|---|---|---|
| Playground | 22 | 4190.2659 |
|  | 27 | 1784.2595 |
|  | 32 | 877.7816 |
|  | 37 | 469.7524 |
| Table | 22 | 2187.9218 |
|  | 27 | 1087.1641 |
|  | 32 | 618.5536 |
|  | 37 | 379.0299 |
| Hallway | 22 | 1092.4658 |
|  | 27 | 290.2119 |
|  | 32 | 311.6206 |
|  | 37 | 182.0746 |
| Tree | 22 | 3141.0672 |
|  | 27 | 1022.0334 |
|  | 32 | 843.1776 |
|  | 37 | 488.2736 |

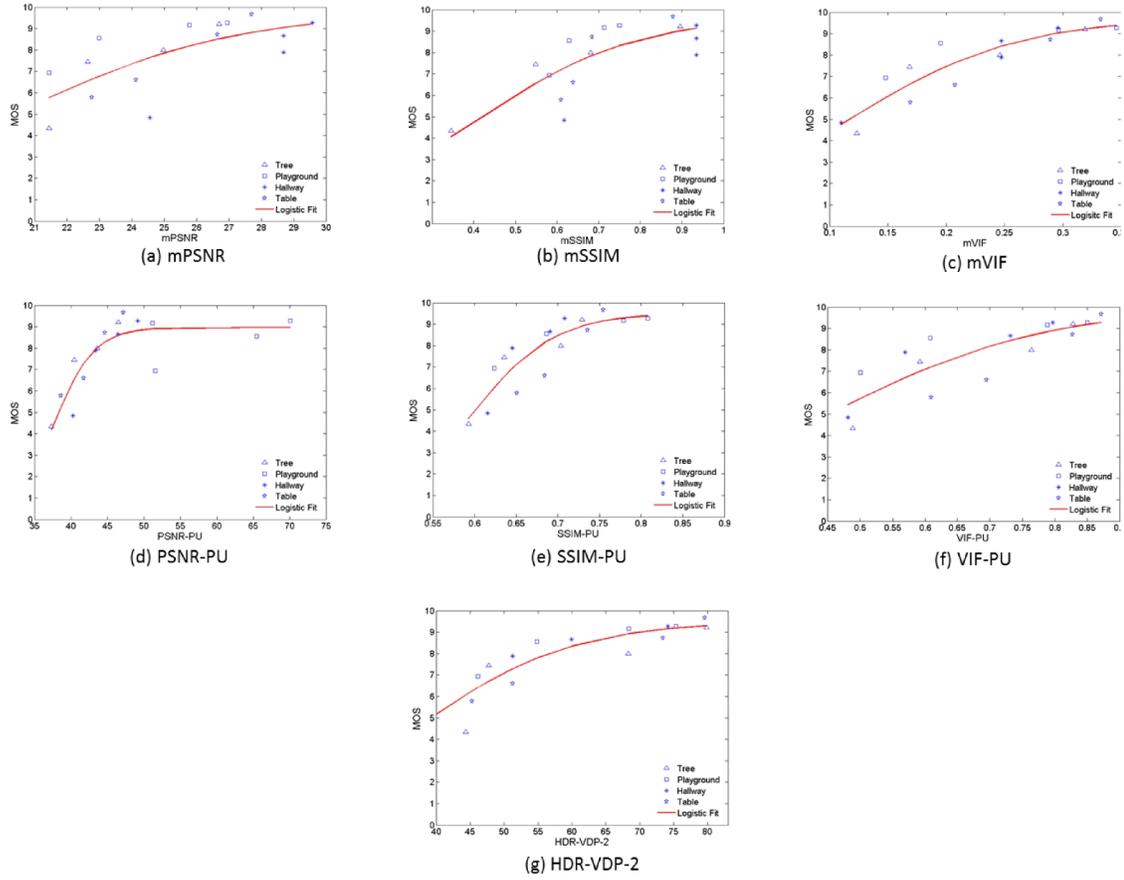

Fig. 5 Subjective results versus objective measure in the presence of compression artifacts: (a) MultiExposure PSNR, (b) MultiExposure SSIM, (c) MultiExposure VIF, (d) PSNR (PU encoding), (e) SSIM (PU encoding), (f) VIF (PU encoding), and (g) HDR-VDP-2

Table II summarizes the results of the correlation between the objective quality scores and the ones of the subjective tests. In order to estimate each metric's accuracy, the Pearson Linear Correlation Coefficient (PCC) and Root Mean Square Error (RMSE) are calculated between MOS values and the obtained objective quality indices. The Spearman Rank Order Correlation Coefficient (SCC) is also computed to estimate the monotonicity in the metrics' results. The PCC and SCC in each column are calculated over the entire video data set. The results are reported based on three impairments categories: a) compression artifacts, b) AWGN, intensity shifting, salt & pepper noise, and low pass filtering, and c) all the impairments used in our study.

As it is observed from Table II, in the presence of the AWGN, intensity shifting, salt & pepper noise, and low pass filtering distortions, VIF with PU encoding yields the best performance compared to other used LDR metrics. However, in the presence of the compression artifacts, HDR-VDP-2 and MultiExposure VIF outperform all other tested metrics. Overall, in the presence of the tested distortions, VIF with PU encoding shows the best performance in predicting the quality of the HDR videos compared to other tested metrics.

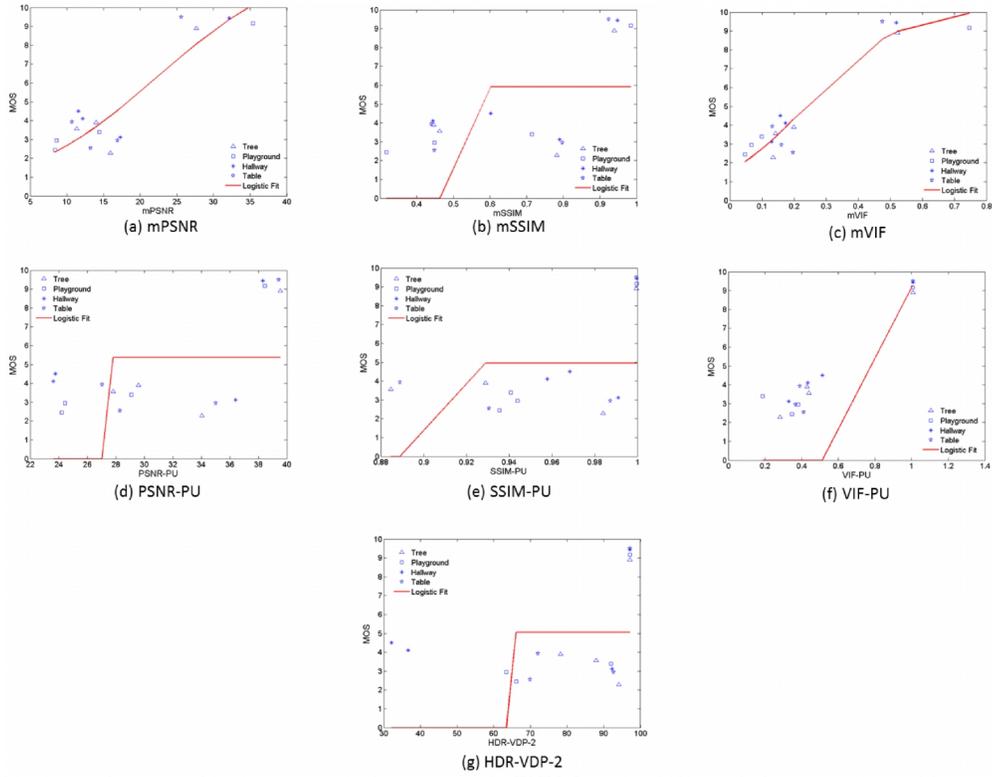

Fig. 6 Subjective results versus objective measure in the presence of AWGN, Intensity shifting, salt & pepper noise, and low pass filtering impairments: (a) MultiExposure PSNR, (b) MultiExposure SSIM, (c) MultiExposure VIF, (d) PSNR (PU encoding), (e) SSIM (PU encoding), (f) VIF (PU encoding), and (g) HDR-VDP-2

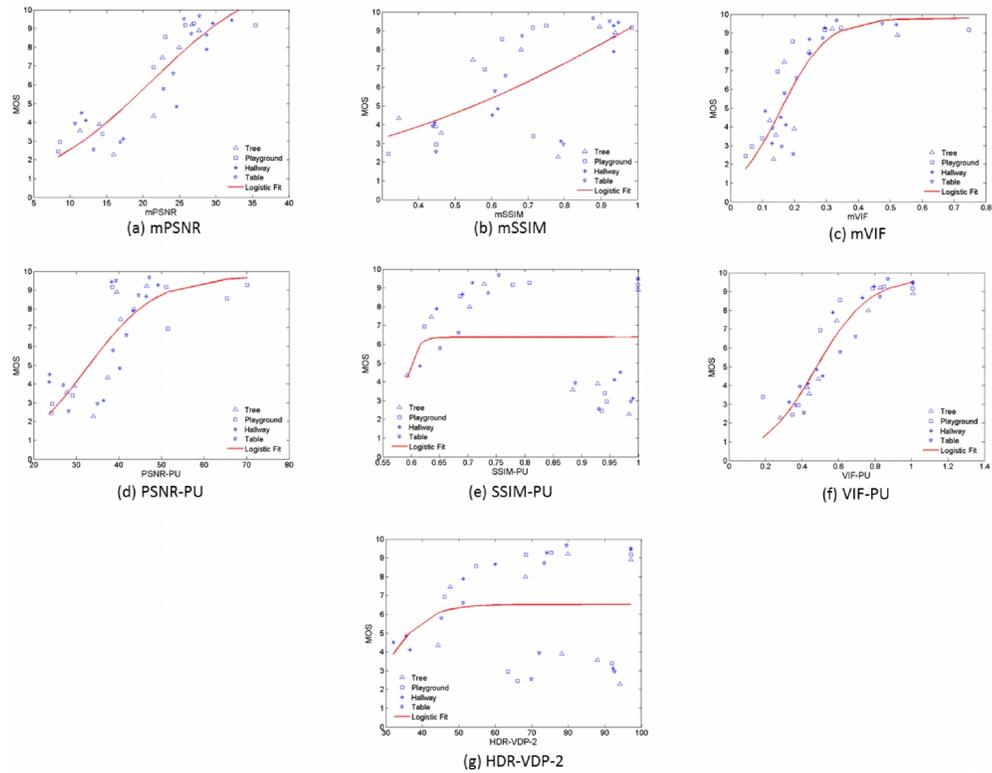

Fig. 7 Subjective results versus objective measure in the presence of AWGN, Intensity shifting, salt & pepper noise, low pass filtering, and compression artifacts: (a) MultiExposure PSNR, (b) MultiExposure SSIM, (c) MultiExposure VIF, (d) PSNR (PU encoding), (e) SSIM (PU encoding), (f) VIF (PU encoding), and (g) HDR-VDP-2

TABLE II
CORRELATION OF SUBJECTIVE RESPONSES WITH PREDICTION OF OBJECTIVE QUALITY METRICS

| Metric/Method | Impairments: AWGN, Intensity shifting, salt & pepper noise, and low pass filtering | | | Impairment: Different levels of compression, QP: 22, 27, 32, 37 | | | Impairments: AWGN, Intensity shifting, salt & pepper noise, low pass filtering, and compression | | |
|---|---|---|---|---|---|---|---|---|---|
| | *Pearson Correlation* | *Spearman Correlation* | *RMSE* | *Pearson Correlation* | *Spearman Correlation* | *RMSE* | *Pearson Correlation* | *Spearman Correlation* | *RMSE* |
| HDR-VDP-2 | 0.3895 | 0.3650 | 9.5679 | 0.8723 | **0.9581** | 0.4493 | 0.1095 | 0.1821 | 6.6816 |
| PSNR (PU encoding) | 0.6893 | 0.3812 | 10.3556 | 0.5717 | 0.7358 | 0.6269 | 0.7354 | 0.7379 | 2.4702 |
| SSIM (PU encoding) | 0.5715 | 0.4857 | 8.5903 | 0.8277 | 0.8992 | 0.5630 | 0.3506 | 0.1230 | 6.8648 |
| VIF (PU encoding) | **0.9731** | **0.8492** | 8.5569 | 0.8334 | 0.8904 | 0.7644 | **0.9163** | **0.9482** | **0.7109** |
| PSNR (MultiExposure) | 0.8886 | 0.4989 | 1.2921 | 0.6849 | 0.7349 | 1.3093 | 0.8846 | 0.8612 | 1.3700 |
| SSIM (MultiExposure) | 0.7231 | 0.4739 | 10.2767 | 0.7410 | 0.7599 | 1.0117 | 0.6732 | 0.6293 | 3.8509 |
| VIF (MultiExposure) | 0.9311 | 0.6961 | **0.5501** | **0.8985** | 0.9264 | **0.3983** | 0.7306 | 0.8456 | 1.7023 |

# 5 Conclusions

The main purpose of the submission was to investigate the performance of existing quality metrics in evaluating the quality of HDR content. To this end five types of distortions (AWGN, intensity shifting, salt & pepper noise, low pass filtering, and compression) were applied on four video sequences of DML-HDR dataset. Experiments results showed that in the presence of compression distortions, HDR-VDP2 and VIF with MultiExposure method outperform all other metrics. Overall VIF using PU encoding yields the best performance in the presence of all the tested impairments.

It is recommended to consider our findings along with the other subjective HDR quality evaluations for choosing an appropriate objective metric for HDR video content.

# 6 Acknowledgement

We would like to thank Dolby Vancouver for donating the HDR prototype display system to Digital Multimedia Lab (UBC), which made performing the HDR video subjective tests possible.